\documentclass[twocolumn,aps,showpacs,pra]{revtex4}
\usepackage{epsfig}
\usepackage[english]{babel}
\usepackage{latexsym}
\usepackage{graphics}
\usepackage{subfigure}
\usepackage{epsfig}
\usepackage{graphicx}
\usepackage{dcolumn}
\usepackage{amsmath}

\begin{document}
\title{Charge-Resonance Effect on Harmonic Generation by Symmetric  Diatomic Molecular Ions in Intense Laser Fields
}
\author{Yanjun Chen$^{1,2}$, Jing Chen$^{1}$, and Jie Liu$^{1*}$}

\date{\today}

\begin{abstract}
In present paper we develop an analytic theory for the harmonic
generation of symmetric diatomic molecular ions beyond two-level
model, emphasizing the influence of charge-resonance (CR) states
those are strongly coupled to electromagnetic fields for large
internuclear distance. With taking  into account  the continuum
states that is ignored in the two-level model and become important
for intense laser case, our model is capable to produce spectrum
for the whole range of harmonic orders consisting of a molecular
plateau due to the CR transition and an atomic-like plateau for a
long-wavelength excitation, and in good agreement with numerical
results from directly solution of the Schr\"odinger equation. Our
theory also identifies the crucial role of the CR states in  the
fine structure of harmonic spectrum and shows that the harmonic
generation in molecular system can be effectively controlled by
adjusting the  internuclear distance.
\end{abstract}
\affiliation{1.Institute of Applied Physics and Computational
Mathematics, P.O.Box 100088, Beijing, P. R. China\\
2.Graduate School, China Academy of Engineering Physics, P.O. Box
8009-30, Beijing, 100088, P. R. China} 
\maketitle

\section{Introduction}
In recent years, high order harmonic generation (HOHG) from atom
and molecule has been the subject of numerous experimental and
theoretical studies, mainly because the emission of high-order
harmonics is a promising method to produce coherent x rays and
attosecond pulses, and an effective ways to study the internal
construction of atom or
molecule\cite{1,2,3,4,5,6,7,8,9,10,11,12,13,14,15,16,17,18,19,20,21,22}.

Compared to atom case, the HOHG spectrum  of molecule systems
demonstrates  novel  properties, e.g., peak splitting and sideband
peaks, Rabi or Mollow triplets effect, and combination of the
atomic-like and molecular-like plateau, molecular alignment
effect, and  two-center interference, to name only a
few\cite{23,24,25,26}. Symmetric molecular system such as $H_2^+$
have pairs of electronic states known as the charge-resonance (CR)
states, (i.e.,$1\sigma_g$,$1\sigma_\mu$.) It is known that the CR
states are strongly coupled to electromagnetic fields at large
internuclear separation $R$. Bandrauk and co-worker first pointed
out the importance of these CR states as sources of highly
nonlinear laser-induced effects in molecules\cite{23}. Ivanov,
Corkum examined the possibility of using the CR states to produce
and coherently control HOHG\cite{24}. Zuo, Chelkowski, and
Bandrauk have showed that symmetric molecular ions in general
produce more efficient harmonic generation than atoms, especially
at large $R$, due to these CR states\cite{25,26}. However, in the
all above discussions, the  theoretical analysis is based on a
two-level model where the continuum (ionization) states are
ignored completely. Since the ground-continuum coupling and
ionization become important for the strong laser fields of
$10^{14}W/cm^2$ and  above, a fully understanding of  the whole
structure of HOHG and the effect of  CR states in  molecule
systems requires to extend to consider the continuum states and
ionization process.

In present paper we develop an analytic theory on the harmonic
generation of symmetric diatomic molecular ions beyond the
two-level model. Our model is capable to  produce  harmonic
structure for the whole range of harmonic order, including the
molecular plateau due to CR transition and the atomic-like plateau
for a long-wavelength excitation, and  agrees well with the
numerical results from directly solving the Schr\"odinger
equation. Our theory identifies the role of the CR states in  the
fine structure of harmonic spectrum and shows that the harmonic
generation in molecular system can be effectively controlled
through CR states by adjusting the internuclear distance.

Our paper is organized as follows. In Sec.II we present our
analytical theory. Our analysis is divided into two cases:
near-resonance region of intermediate internuclear distance and
the strong-coupling region of large internuclear distance. We
will derive the time-dependent amplitudes of the ground, excited
and continuum states, then calculate dipolar moments and their
Fourier transformation. The analytic expressions of the amplitudes
of  HOHG for the whole range of harmonic order is given in this
section. Sec.III presents our numerical results. Our theory is
applied to 1D symmetric diatomic molecule model, making analysis
on the structure of HOHG and compare with the results from
directly solving Schr\"odinger equation. Sec.V is our conclusion.

\section{Analytic theory}
We consider a diatomic molecule ion in single-electron
approximation under the influence of a linear polarized laser
field $\mathbf{A}(t)=(E\cos({\omega}t),0,0)$, where $E{\omega}$ is
the absolute amplitude of the external electric field and
${\omega}$ is the frequency of the external field. The Hamiltonian
of the model molecule studied here is
\begin{equation*}
\begin{split}
H(t)=(\mathbf{p}-\mathbf{A}(t))^{2}/2+V(\mathbf{r}),
\end{split}
\end{equation*}
where $\mathbf{p}$ is the canonical momentum and $V(\mathbf{r})$
is the binding potential.

 Under strong-field conditions
\cite{27,28}, it is reasonable to assume that, (a) Except the
ground state $|0\rangle$ and the first excited state$|1\rangle$,
the contribution from  other bound states
 can be neglected;(b) The depletion of the ground
state and the first excited state is small;(c) In the continuum,
the electron can be treated as a free particle moving in the
electric field without considering Coulomb potential. Then, the
time-dependent wave functions can be expanded as
\begin{equation}
|\psi(t)\rangle=e^{-iE_{0}t}[a(t)|0\rangle+b(t)|1\rangle+\int
d\mathbf{p}c_{\mathbf{p}}(t)|\mathbf{p}\rangle],
\end{equation}
where $-E_0$ is the ionization potential of the ground state,
$a(t)$ is the ground-state amplitude, $b(t)$ is the first excited
state amplitude and $c_{\mathbf{p}}(t)$ are the amplitudes of the
corresponding continuum states. While the ionization is weak, by
neglecting the depletion of the ground and the first excited state
and setting $|a(t)|^{2}+|b(t)|^{2}=1$, the formulation of $a(t)$
and $b(t)$ can be obtained by a two-level approximation. We have
factor out here free oscillations of the ground-state amplitude
by the bare frequency $-E_{0}$. The Schr$\ddot o$dinger equation
for $c_{\mathbf{p}}(t)$ reads as
\begin{equation}
\dot{c}_{\mathbf{p}}(t)=i
\mathbf{A}(t)\mathbf{p}[a(t)\langle\mathbf{p}|0\rangle+b(t)\langle\mathbf{p}|1\rangle]+
[(\mathbf{p}-\mathbf{A}(t))^{2}/2-E_{0}]c_{\mathbf{p}}(t).
\end{equation}
(2) can be solved exactly and $c_{\mathbf{p}}(t)$ can be written
in the closed form
\begin{eqnarray}
\begin{split}
c_{\mathbf{p}}(t)=i\int_0^t
dt'\mathbf{A}(t')\mathbf{p}[a(t')\langle\mathbf{p}|0\rangle+b(t')\langle\mathbf{p}|1\rangle]\\
\times e^{-i\int_{t'}^{t}
[(\mathbf{p}-\mathbf{A}(t''))^{2}/2-E_{0}]dt''}.
\end{split}
\end{eqnarray}
Using Eq. (1) and (3), the mechanical momentum
$\mathbf{p}-\mathbf{A}(t)$ component of the time-dependent dipole
moment is
\begin{equation}
{\bf
D}(t)=\langle\psi(t)|{\mathbf{p}}-\mathbf{A}(t)|\psi(t)\rangle.
\end{equation}
Neglecting the term
$\langle\psi(t)|\mathbf{A}(t)|\psi(t)\rangle$=$\mathbf{A}(t)$,
which only includes the fundamental frequency ${\omega}$, and the
contribution from C-C part and considering only the transitions
back to the ground state and the first excited state, we obtain
${\bf D}(t)=D_{0}(t)+D_{1}(t)$,where
\begin{gather}
\begin{split}
D_{0}(t)=i\int d\mathbf{p}\int_0^t
dt'a^{*}(t)\langle0|{\mathbf{p}}|\mathbf{p}\rangle\mathbf{A}(t')\cdot\mathbf{p}\big[a(t')\langle\mathbf{p}|0\rangle\\
+b(t')\langle\mathbf{p}|1\rangle\big] \times
e^{-iS(\mathbf{p},t,t')},
\end{split}\\
\begin{split}
D_{1}(t)=i\int d\mathbf{p}\int_0^t
dt'b^{*}(t)\langle1|{\mathbf{p}}|\mathbf{p}\rangle\mathbf{A}(t')\cdot\mathbf{p}\big[a(t')\langle\mathbf{p}|0\rangle\\
+b(t')\langle\mathbf{p}|1\rangle\big] \times
e^{-iS(\mathbf{p},t,t')},
\end{split}
\end{gather}
where $S(\mathbf{p},t,t')=\int_{t'}^{t}
\left[(\mathbf{p}-\mathbf{A}(t''))^{2}/2-E_{0}\right]dt''$.
$D_{0}(t)$ denotes the transition back to the ground state, and
$D_{1}(t)$  denotes the transition back to the first excited
state. Each of $D_{0}(t)$ and $D_{1}(t)$ includes two different
terms, which denote two different moments, respectively. For
$D_{0}(t)$, they are
$|0\rangle\rightarrow|\mathbf{p}\rangle\rightarrow|0\rangle$
(denoted by $D_{0}^{0}(t)$) and
$|1\rangle\rightarrow|\mathbf{p}\rangle\rightarrow|0\rangle$
(denoted by $D_{0}^{1}(t)$), which denote the transition from the
ground state to the continuum state, then back to the ground
state($D_{0}^{0}(t)$), or from the first excited state to the
continuum state, then back to the ground state($D_{0}^{1}(t)$);
for $D_{1}(t)$,they are
$|0\rangle\rightarrow|\mathbf{p}\rangle\rightarrow|1\rangle$
(denoted by $D_{1}^{0}(t)$) and
$|1\rangle\rightarrow|\mathbf{p}\rangle\rightarrow|1\rangle$
(denoted by $D_{1}^{1}(t)$), which denote the transition from the
ground state to the continuum state, then back to first excited
state($D_{1}^{0}(t)$), or from the first excited state to the
continuum state, then back to first excited state($D_{1}^{1}(t)$).
We should discuss these moments at intermediate $R$ and large $R$,
respectively.

\subsection{Intermediate $R$ with $(E_{1}-E_{0})/{\omega}\simeq 1$}
For a two-level system, the time-dependent wave function can be
written as $|\psi(t)\rangle$=$a(t)|0\rangle+b(t)|1\rangle$, where
$|0\rangle$ and $|1\rangle$ are the ground-state and the first
excited state of the unperturbed Hamiltonian
$H_{0}={{\mathbf{p}}}^{2}+V({\vec{r}})$ with
$H_{0}|0\rangle=E_{0}|0\rangle$ and
$H_{0}|1\rangle=E_{1}|1\rangle$. While
$(E_{1}-E_{0})/{\omega}\simeq 1$, in the rotating-wave
approximation, the solution for $a(t)$ and $b(t)$ can be written
as
\begin{eqnarray*}
\left\{
\begin{array}{ll}
a(t)=e^{-iE_0 t-i\int_0^t [\mathbf{A}^{2}(t')/2]dt'}(a_1 e ^{iQt/2}+a_2 e^{-iQt/2})e^{-i\xi t/2},\\
b(t)=e^{-iE_1 t-i\int_0^t [\mathbf{A}^{2}(t')/2]dt'}(b_1
e^{iQt/2}+b_2 e^{-iQt/2})e^{i\xi t/2},
\end{array}
\right.
\end{eqnarray*}
and by the Bessel function formula\cite{29},we obtain finally
\begin{eqnarray}
 \left\{
\begin{array}{ll}
a(t)=\sum_{n=-\infty}^{n=+\infty}J_{n}(E^{2}/8{\omega})(a_{1}e^{-iA_{1}t}+a_{2}e^{-iA_{2}t})\\
b(t)=\sum_{n=-\infty}^{n=+\infty}J_{n}(E^{2}/8{\omega})(b_{1}e^{-iB_{1}t}+b_{2}e^{-iB_{2}t}),
\end{array}
\right.
\end{eqnarray}
where $E{\omega}$ is the absolute amplitude of the external
electric field and$A_{1}=E^{2}/4+2n{\omega}-Q/2+\xi/2,$
$A_{2}=E^{2}/4+2n{\omega}+Q/2+\xi/2,$
$B_{1}=E_{1}+E^{2}/4+2n{\omega}-Q/2-\xi/2-E_{0},$
$B_{2}=E_{1}+E^{2}/4+2n{\omega}+Q/2-\xi/2-E_{0},$ where
$\xi$=$E_1-E_0-{\omega}$; $Q$=$\sqrt{|Q_{R}|^{2}+\xi^{2}}$,
$Q_{R}$=$ED$, $D$=$\langle0|{\mathbf{p}}|1\rangle$ is the matrix
element of the electric dipole moment, and
 $a_{1},a_{2},b_{1}$ and
$b_{2}$ are constants of integration which are determined from the
initial conditions
$|\psi(0)\rangle$=$a(0)|0\rangle+b(0)|1\rangle$:
\begin{eqnarray}
\left\{
\begin{array}{ll}
a_{1}=[(Q+\xi)a(0)+Q_{R}b(0)]/(2Q)\\
a_{2}=[(Q-\xi)a(0)-Q_{R}b(0)]/(2Q)\\
b_{1 }=[(Q-\xi)b(0)+Q_{R}^{*}a(0)]/(2Q)\\ b_{2}=
[(Q+\xi)b(0)-Q_{R}^{*}a(0)]/(2Q),
\end{array}
\right.
\end{eqnarray}
where $Q_{R}^{*}$ is the conjugated element of $Q_{R}$. For the
convenience of latter calculation, we have multiplied $a(t)$ and
$b(t)$ by a phasic factor $e^{iE_{0}t}$. Substituting (8) into (5)
and (6), then it follows:
\subsubsection{$|0\rangle\rightarrow|\mathbf{p}\rangle\rightarrow|0\rangle$}
For $D_{0}(t)$, above all we consider the
moment$|0\rangle\rightarrow|\mathbf{p}\rangle\rightarrow|0\rangle$
, i.e.,
\begin{equation}
\begin{split}
D_{0}^{0}(t)=i\int d\mathbf{p}\int_0^t
dt'a^{*}(t)\langle0|{\mathbf{p}}|\mathbf{p}\rangle{\mathbf{A}(t')\cdot\mathbf{p}a(t')\langle\mathbf{p}|0\rangle
e^{-iS(\mathbf{p},t,t')}},
\end{split}
\end{equation}
For $e^{-iS(\mathbf{p},t,t')}$, by the formula $e^{-i\gamma
sin(2{\omega}t)+i\kappa
Esin({\omega}t)}=\sum_{n=-\infty}^{n=+\infty}
J_{n}(-\kappa,\gamma)e^{-in{\omega}t}$, we can obtain
\begin{equation}
\begin{split}
e^{-iS(\mathbf{p},t,t')}=& e^{-i(E^{2}/4+\mathbf{p}^{2}/2-E_{0})(t-t')}\sum_{n,m=-\infty}^{n,m=+\infty},\\
&
J_{n}(\frac{-p_{x}E}{{\omega}},\frac{E^{2}}{8{\omega}})J_{m}(\frac{p_{x}E}{{\omega}},\frac{-E^{2}}{8{\omega}})
e^{-in{\omega}t}e^{-im{\omega}t'}\nonumber,
\end{split}
\end{equation}
substituting it into (9), and integrating over $t'$, we obtain
\begin{equation}
\begin{split}
& D_{0}^{0}(t) = i\int d\mathbf{p} \frac{E}{2}\langle
0|\mathbf{p}\rangle \mathbf{p}(\mathbf{p}_{x})\langle
\mathbf{p}|0\rangle\\
&
\sum_{n,m,n_{1},m_{1}=-\infty}^{n,m,n_{1},m_{1}=+\infty}\bigg\{\frac{a_{1}^{*}a_{1}e^{-i(C_{1}+A_{1}'+D_{1}-A_{1})t}}
{-i(C_{1}+A_{1}')}\nonumber\\
&  +\frac{a_{1}^{*}a_{2}e^{-i(C_{1}+A_{2}'+D_{1}-A_{1})t}}
{-i(C_{1}+A_{2}')}+\frac{a_{1}^{*}a_{1}e^{-i(C_{2}+A_{1}'+D_{1}-A_{1})t}}
{-i(C_{2}+A_{1}')}\nonumber\\
&  +\frac{a_{1}^{*}a_{2}e^{-i(C_{2}+A_{2}'+D_{1}-A_{1})t}}
{-i(C_{2}+A_{2}')}+\frac{a_{2}^{*}a_{1}e^{-i(C_{1}+A_{1}'+D_{1}-A_{2})t}}
{-i(C_{1}+A_{1}')}\nonumber\\
&  +\frac{a_{2}^{*}a_{2}e^{-i(C_{1}+A_{2}'+D_{1}-A_{2})t}}
{-i(C_{1}+A_{2}')}+\frac{a_{2}^{*}a_{1}e^{-i(C_{2}+A_{1}'+D_{1}-A_{2})t}}
{-i(C_{2}+A_{1}')}\nonumber\\
&  +\frac{a_{2}^{*}a_{2}e^{-i(C_{2}+A_{2}'+D_{1}-A_{2})t}}
{-i(C_{2}+A_{2}')}\bigg\}J_{n}J_{m}J_{n_{1}}J_{m_{1}}\nonumber,
\end{split}
\end{equation}
where
$J_{m}$=$J_{m}(\frac{p_{x}E}{{\omega}},\frac{-E^{2}}{8{\omega}})$,
$J_{m_{1}}$=$J_{m_{1}}(\frac{-p_{x}E}{{\omega}},\frac{E^{2}}{8{\omega}})$,
$J_{n}$=$J_{n}(\frac{E^{2}}{8{\omega}})$,
$J_{n_{1}}$=$J_{n_{1}}(\frac{E^{2}}{8{\omega}})$,and $A_{1}=
E^{2}/4+2n_{1}{\omega}-Q/2+\xi/2,$
$A_{2}=E^{2}/4+2n_{1}{\omega}+Q/2+\xi/2,$
$C_{1}=m{\omega}-\mathbf{p}^{2}/2-E^{2}/4+E_{0}-{\omega},$
$C_{2}=m{\omega}-\mathbf{p}^{2}/2-E^{2}/4+E_{0}+{\omega},$
$A_{1}'=E^{2}/4+2n{\omega}-Q/2+\xi/2,$
$A_{2}'=E^{2}/4+2n{\omega}+Q/2+\xi/2,$$D_{1}=\mathbf{p}^{2}/2+E^{2}/4-E_{0}+m_{1}{\omega}.$

The Fourier component of $D_{0}^{0}(t)$ reads as
\begin{equation}
\begin{split}
&P_{0}^{0}({\omega}') = -i\int d\mathbf{p}E\pi\langle
0|\mathbf{p}\rangle \mathbf{p}p_{x}\langle
\mathbf{p}|0\rangle\sum_{n,m,n_{1},m_{1}=-\infty}^{n,m,n_{1},m_{1}=+\infty}J_{n}J_{m}\\
&\Bigg\{\frac{a_{1}^{*}a_{1}\delta\left[(m+m_{1}-1+2(n-n_{1})){\omega}-{\omega}'\right]}{(C_{1}+A_{1}')}\\
&+\frac{a_{2}^{*}a_{2}\delta\left[(m+m_{1}-1+2(n-n_{1})){\omega}-{\omega}'\right]}
{(C_{1}+A_{2}')}\\
&+\frac{a_{1}^{*}a_{1}\delta\left[(m+m_{1}+1+2(n-n_{1})){\omega}-{\omega}'\right]}
{(C_{2}+A_{1}')}\\
&+\frac{a_{2}^{*}a_{2}\delta\left[(m+m_{1}+1+2(n-n_{1})){\omega}-{\omega}'\right]}
{(C_{2}+A_{2}')}\\
&+\frac{a_{1}^{*}a_{2}\delta\left[(m+m_{1}-1+2n-2n_{1}){\omega}+Q-{\omega}'\right]}
{(C_{1}+A_{2}')}\\
&+\frac{a_{1}^{*}a_{2}\delta\left[(m+m_{1}+1+2n-2n_{1}){\omega}+Q-{\omega}'\right]}
{(C_{2}+A_{2}')}\\
&+\frac{a_{2}^{*}a_{1}\delta\left[(m+m_{1}-1+2n-2n_{1}){\omega}-Q-{\omega}'\right]}
{(C_{1}+A_{1}')}\\
&+\frac{a_{2}^{*}a_{1}\delta\left[(m+m_{1}+1+2n-2n_{1}){\omega}-Q-{\omega}'\right]}
{(C_{2}+A_{1}')}\Bigg\}J_{n_{1}}J_{m_{1}}
\end{split}
\end{equation}

For(10),the parity of $\langle
0|\mathbf{p}\rangle\mathbf{p}p_{x}\langle \mathbf{p}|0\rangle$ for
the integral over $\mathbf{p}$ is even, according to the property
of Bessel function, only when $m+m_{1}$ is even, the value of Eq.
(10) for the integral over $\mathbf{p}$ is not zero. Fourier
components on the right-hand side of (10) can be divided into two
parts, odd Fourier components of $(m+m_{1}+2n-2n_{1}\pm
1){\omega}$, and no-integer Fourier components of
$(m+m_{1}+2n-2n_{1}\pm 1){\omega}\pm Q$ .It also can be seen from
(10), the Rabi oscillation of the ground and first excited state
should not contribute to the parity of integer harmonic, but
induces the symmetrical splitting of odd harmonic. The splitting
separations around each odd harmonic all take the same value $Q$.
If $Q/{\omega}$ is integer of odd number, then the odd harmonic
sidebands should coincide at the even harmonic ($2n{\omega}$)
position, thus giving rise to radiation of even harmonics.
\subsubsection{$|1\rangle\rightarrow|\mathbf{p}\rangle\rightarrow|0\rangle$}
For $|1\rangle\rightarrow|\mathbf{p}\rangle\rightarrow|0\rangle$,
i.e.,
\begin{equation}
D_{0}^{1}(t)=i\int d\mathbf{p}\int_0^t
dt'a^{*}(t)\langle0|{\mathbf{p}}|\mathbf{p}\rangle{\mathbf{A}(t')\cdot\mathbf{p}b(t')\langle\mathbf{p}|1\rangle
e^{-is(\mathbf{p},t,t')}}\nonumber.
\end{equation}
Analogous to
$|0\rangle\rightarrow|\mathbf{p}\rangle\rightarrow|0\rangle$, the
Fourier component of $D_{0}^{1}(t)$ reads as
\begin{equation}
\begin{split}
& P_{0}^{1}({\omega}')=-i\int d\mathbf{p}E\pi\langle
0|\mathbf{p}\rangle \mathbf{p}p_{x}\langle
\mathbf{p}|1\rangle\sum_{n,m,n_{1},m_{1}=-\infty}^{n,m,n_{1},m_{1}=+\infty}J_{n}J_{m}\\
&
\Bigg\{\frac{a_{1}^{*}b_{1}\delta\left[(m+m_{1}+2n-2n_{1}){\omega}-{\omega}'\right]}
{(C_{1}+B_{1}')}\\
&+\frac{a_{2}^{*}b_{2}\delta\left[(m+m_{1}+2n-2n_{1}){\omega}-{\omega}'\right]}
{(C_{1}+B_{2}')}\\
&
+\frac{a_{1}^{*}b_{1}\delta\left[(m+m_{1}+2+2n-2n_{1}){\omega}-{\omega}'\right]}
{(C_{2}+B_{1}')}\\
&+\frac{a_{2}^{*}b_{2}\delta\left[(m+m_{1}+2+2n-2n_{1}){\omega}-{\omega}'\right]}
{(C_{2}+B_{2}')}\\
&
+\frac{a_{1}^{*}b_{2}\delta\left[(m+m_{1}+2n-2n_{1}){\omega}+Q-{\omega}'\right]}
{(C_{1}+B_{2}')}\\
&
+\frac{a_{1}^{*}b_{2}\delta\left[(m+m_{1}+2+2n-2n_{1}){\omega}+Q-{\omega}'\right]}
{(C_{2}+B_{2}')}\\
&
+\frac{a_{2}^{*}b_{1}\delta\left[(m+m_{1}+2n-2n_{1}){\omega}-Q-{\omega}'\right]}
{(C_{1}+B_{1}')}\\
&
+\frac{a_{2}^{*}b_{1}\delta\left[(m+m_{1}+2+2n-2n_{1}){\omega}-Q-{\omega}'\right]}
{(C_{2}+B_{1}')}\Bigg\}J_{n_{1}}J_{m_{1}},
\end{split}
\end{equation}
where  the definitions of $C_{1}$ and $C_{2}$ are the same as in
$|0\rangle\rightarrow|\mathbf{p}\rangle\rightarrow|0\rangle$, and
$B_{1}'=E_{1}+E^{2}/4+2n{\omega}-Q/2-\xi/2-E_{0},$
$B_{2}'=E_{1}+E^{2}/4+2n{\omega}+Q/2-\xi/2-E_{0}.$

 For(11), the
parity of $\langle 0|\mathbf{p}\rangle\mathbf{p}p_{x}\langle
\mathbf{p}|1\rangle$ for the integral over  $\mathbf{p}$ is odd,
only when $m+m_{1}$ is odd, the value of(11) for the integral over
$\mathbf{p}$  is not zero. Fourier components on the right-hand
side of (11) can also be divided into two parts, odd Fourier
components of $(m+m_{1}+2n-2n_{1}){\omega}$ and
$(m+m_{1}+2n-2n_{1}+2){\omega}$, and no-integer Fourier components
of $(m+m_{1}+2n-2n_{1}){\omega}\pm Q$ and
$(m+m_{1}+2+2n-2n_{1}){\omega}\pm Q$.
\subsubsection{$|0\rangle\rightarrow|\mathbf{p}\rangle\rightarrow|1\rangle$}
For $|0\rangle\rightarrow|\mathbf{p}\rangle\rightarrow|1\rangle$,
i.e.,
\begin{equation}
D_{1}^{0}(t)=i\int d\mathbf{p}\int_0^t
dt'b^{*}(t)\langle1|{\mathbf{p}}|\mathbf{p}\rangle{\mathbf{A}(t')\cdot\mathbf{p}a(t')\langle\mathbf{p}|0\rangle
e^{-iS(\mathbf{p},t,t')}}.\nonumber
\end{equation}
The Fourier component of $D_{1}^{0}(t)$ reads as
\begin{equation}
\begin{split}
&P_{1}^{0}({\omega}') = -i\int d\mathbf{p}E\pi\langle
1|\mathbf{p}\rangle \mathbf{p}p_{x}\langle
\mathbf{p}|0\rangle\sum_{n,m,n_{1},m_{1}=-\infty}^{n,m,n_{1},m_{1}=+\infty}J_{n}J_{m}\\
&
\Bigg\{\frac{b_{1}^{*}a_{1}\delta\left[(m+m_{1}-2+2n-2n_{1}){\omega}-{\omega}'\right]}
{(C_{1}+A_{1}')}\\
&+\frac{b_{2}^{*}a_{2}\delta\left[(m+m_{1}-2+2n-2n_{1}){\omega}-{\omega}'\right]}
{(C_{1}+A_{2}')}\\
&
+\frac{b_{1}^{*}a_{1}\delta\left[(m+m_{1}+2n-2n_{1}){\omega}-{\omega}'\right]}
{(C_{2}+A_{1}')}\\
&+\frac{b_{2}^{*}a_{2}\delta\left[(m+m_{1}+2n-2n_{1}){\omega}-{\omega}'\right]}
{(C_{2}+A_{2}')}\\
&
+\frac{b_{1}^{*}a_{2}\delta\left[(m+m_{1}-2+2n-2n_{1}){\omega}+Q-{\omega}'\right]}
{(C_{1}+A_{2}')}\\
&
+\frac{b_{1}^{*}a_{2}\delta\left[(m+m_{1}+2n-2n_{1}){\omega}+Q-{\omega}'\right]}
{(C_{2}+A_{2}')}\\
&
+\frac{b_{2}^{*}a_{1}\delta\left[(m+m_{1}-2+2n-2n_{1}){\omega}-Q-{\omega}'\right]}
{(C_{1}+A_{1}')}\\
&
+\frac{b_{2}^{*}a_{1}\delta\left[(m+m_{1}+2n-2n_{1}){\omega}-Q-{\omega}'\right]}
{(C_{2}+A_{1}')}\Bigg\}J_{n_{1}}J_{m_{1}}.
\end{split}
\end{equation}
where the definitions of $C_{1}$, $C_{2}$, $A_{1}'$, $A_{2}'$, are
the same as in
$|0\rangle\rightarrow|\mathbf{p}\rangle\rightarrow|0\rangle$.

For(12), only when $m+m_{1}$ is odd, the value of(12) for the
integral over $\mathbf{p}$ is not zero. Fourier components on the
right-hand side of (12) also can be divided into two parts, odd
Fourier components of $(m+m_{1}+2n-2n_{1}){\omega}$ and
$(m+m_{1}+2n-2n_{1}-2){\omega}$, and no-integer Fourier components
of $(m+m_{1}+2n-2n_{1}){\omega}\pm Q$ and
$(m+m_{1}-2+2n-2n_{1}){\omega}\pm Q$.
\subsubsection{$|1\rangle\rightarrow|\mathbf{p}\rangle\rightarrow|1\rangle$}
\begin{equation}
D_{1}^{1}(t)=i\int d\mathbf{p}\int_0^t
dt'b^{*}(t)\langle1|{\mathbf{p}}|\mathbf{p}\rangle{\mathbf{A}(t')\cdot\mathbf{p}b(t')\langle\mathbf{p}|1\rangle
e^{-is(\mathbf{p},t,t')}}\nonumber.
\end{equation}
The Fourier component of $D_{1}^{1}(t)$ reads as
\begin{eqnarray}
\begin{split}
& P_{1}^{1}({\omega}') = -i\int d\mathbf{p}E\pi\langle
1|\mathbf{p}\rangle \mathbf{p}p_{x}\langle
\mathbf{p}|1\rangle\sum_{n,m,n_{1},m_{1}=-\infty}^{n,m,n_{1},m_{1}=+\infty}J_{n}J_{m}\\
&
\Bigg\{\frac{b_{1}^{*}b_{1}\delta\left[(m+m_{1}-1+2n-2n_{1}){\omega}-{\omega}'\right]}
{(C_{1}+B_{1}')}\\
&+\frac{b_{2}^{*}b_{2}\delta\left[(m+m_{1}-1+2n-2n_{1}){\omega}-{\omega}'\right]}
{(C_{1}+B_{2}')}\\
&
+\frac{b_{1}^{*}b_{1}\delta\left[(m+m_{1}+1+2n-2n_{1}){\omega}-{\omega}'\right]}
{(C_{2}+B_{1}')}\\
&+\frac{b_{2}^{*}b_{2}\delta\left[(m+m_{1}+1+2n-2n_{1}){\omega}-{\omega}'\right]}
{(C_{2}+B_{2}')}\\
\end{split}
\end{eqnarray}
\begin{eqnarray*}
\begin{split}
&
+\frac{b_{1}^{*}b_{2}\delta\left[(m+m_{1}-1+2n-2n_{1}){\omega}+Q-{\omega}'\right]}
{(C_{1}+B_{2}')}\\
&
+\frac{b_{1}^{*}b_{2}\delta\left[(m+m_{1}+1+2n-2n_{1}){\omega}+Q-{\omega}'\right]}
{(C_{2}+B_{2}')}\\
&
+\frac{b_{2}^{*}b_{1}\delta\left[(m+m_{1}-1+2n-2n_{1}){\omega}-Q-{\omega}'\right]}
{(C_{1}+B_{1}')}\\
&
+\frac{b_{2}^{*}b_{1}\delta\left[(m+m_{1}+1+2n-2n_{1}){\omega}-Q-{\omega}'\right]}
{(C_{2}+B_{1}')}\Bigg\}J_{n_{1}}J_{m_{1}},
\end{split}
\end{eqnarray*}
where the value of $C_{1}$, $C_{2}$, $B_{1}'$ and $B_{2}'$ is the
same as in
$|1\rangle\rightarrow|\mathbf{p}\rangle\rightarrow|0\rangle$.

For(13), the parity of $\langle
1|\mathbf{p}\rangle\mathbf{p}p_{x}\langle \mathbf{p}|1\rangle$ for
the integral over $\mathbf{p}$ is even, only when $m+m_{1}$ is
even, $P_{1}^{1}$ is not zero. Fourier components on the
right-hand side of (13) include odd Fourier components of
$(m+m_{1}+2n-2n_{1}\pm 1){\omega}$, and no-integer Fourier
components of $(m+m_{1}+2n-2n_{1}\pm 1){\omega}\pm Q$.

Using(10)-(13), the Fourier component of $\textbf{D}(t)$ reads as
\begin{eqnarray}
P({\omega}') &=&
\int\limits_{t=-\infty}^{t=+\infty}dt\left[D^{0}_{0}(t)+D^{0}_{1}(t)+D^{1}_{0}(t)+
D^{1}_{1}(t)\right]e^{-i{\omega}'t}\nonumber\\
&=&
P^{0}_{0}({\omega}')+P^{0}_{1}({\omega}')+P^{1}_{0}({\omega}')+
P^{1}_{1}({\omega}').
\end{eqnarray}

It can be seen from (10)-(13), no matter initially the system
being in the ground state or the first excited state or the
coherent superposition state\cite{21}, the right-hand side of (14)
only includes odd harmonic with the symmetrical splitting of $\pm
Q$.

It should be noted that the rotating-wave approximation is
applicable only when $(E_{1}-E_{0})/{\omega}\simeq 1$ and the
external field is weak. However, in case of stronger field where
tunnelling ionization is prominent,  our model at intermediate $R$
is considered to be an approximative approach.  The splitting
separation can also  be calculated by the effective Rabi frequency
or quasienergies of the Flock states of the system\cite{25,26,30}.
But for some specific intermediate $R$, the molecule ions have
relative high ionization rate in the presence of relative weak
field and if the external field frequency ${\omega}$ accords with
$(E_{1}-E_{0})/{\omega}\simeq 1$, our model can give an adequate
description of the process.

When $(E_{1}-E_{0})/{\omega}\simeq0$, according to\cite{25}, the
energy separation of quasienergy (Floquet or dressed) states
around each even harmonic accords with ${\omega}_{q}=\triangle
EJ_{0}(2Q'_{R}/{\omega})\simeq0$, (where $Q'_{R}=ED',
D'=\langle0|\textbf{x}|1\rangle$, $E$ is the absolute amplitude of
the external field here), in other words, when
$R\rightarrow\infty$, all of the large Rabi splittings of the odd
harmonics should converge towards even harmonic\cite{31}. However
it can be showed that in this case the amplitudes of all of even
harmonic should be zero, so at very large $R$, only odd harmonic
can be produced. The point can be made clearer by the latter
analysis of (16)-(19).
\subsection{Large R and $(E_{1}-E_{0})/{\omega}\simeq 0$}
When $(E_{1}-E_{0})/{\omega}\simeq 0$, assuming $E_{1}$=$E_{0}$
and using the Bessel function formula $e^{-i\zeta
sin({\omega}t)}=\sum_{n=-\infty}^{n=+\infty}J_{n}(\zeta)e^{-i{\omega}t}$,
the two-level approximation solution of $a(t)$ and $b(t)$ can be
written as:
\begin{eqnarray}
\left\{
\begin{array}{ll}
\begin{split}
&a(t)=
\frac{1}{2}e^{-i[\frac{E^{2}}{4}t]}\bigg[u(0)\sum\limits_{n=-\infty}^{n=+\infty}J_{n}(\frac{-|Q_R|}{{\omega}},\frac{E^{2}}{8{\omega}})\\
&\times e^{-in{\omega}t}+v(0)\sum\limits_{m=-\infty}^{m=+\infty}J_{m}(\frac{-|Q_R|}{{\omega}},\frac{E^{2}}{8{\omega}})e^{-im{\omega}t}\bigg]\\
&b(t)=
\frac{|Q_R|}{2Q_R}e^{-i[\frac{E^{2}}{4}t]}\bigg[u(0)\sum\limits_{n=-\infty}^{n=+\infty}J_{n}(\frac{-|Q_R|}{{\omega}},\frac{E^{2}}{8{\omega}})\\
&\times e^{-in{\omega}t}-v(0)\sum\limits_{m=-\infty}^{m=+\infty}J_{m}(\frac{-|Q_R|}{{\omega}},\frac{E^{2}}{8{\omega}})e^{-im{\omega}t}\bigg],\\
\end{split}
\end{array}
\right.
\end{eqnarray}
where $u(0)=a(0)+\frac{Q_R}{|Q_R|}b(0)$, and
$v(0)=a(0)-\frac{Q_R}{|Q_R|}b(0)$. For convenience, we have
multiplied $a(t)$ and $b(t)$ by a phasic factor $e^{iE_{0}t}$,
respectively. If initially the system is in the ground state
$|0\rangle$, then $u(0)=1, v(0)=1;$ if initially the system is in
the first excited state $|1\rangle$, then $u(0)=\frac{Q_R}{|Q_R|},
v(0)=-\frac{Q_R}{|Q_R|}$;  if initially the system is in the
coherent superposition state
$\frac{\sqrt{2}}{2}|0\rangle+\frac{\sqrt{2}}{2}|1\rangle$, then
$u(0)=\frac{\sqrt{2}}{2}\left(1+\frac{Q_R}{|Q_R|}\right),$
$v(0)=\frac{\sqrt{2}}{2}\left(1-\frac{Q_R}{|Q_R|}\right).$

 Substituting (15) into (5) and
(6) and considering only the transitions back to the ground state
and the first excited state, one can obtain four different
transition moments
$|0\rangle\rightarrow|\mathbf{p}\rangle\rightarrow|0\rangle$,$|1\rangle\rightarrow|\mathbf{p}\rangle\rightarrow|0\rangle$
,$|0\rangle\rightarrow|\mathbf{p}\rangle\rightarrow|1\rangle$
,$|1\rangle\rightarrow|\mathbf{p}\rangle\rightarrow|1\rangle$.
Accordingly, the Fourier component of the four moments can be
written as:
\begin{eqnarray}
\begin{split}
&P_{0}^{0}({\omega}')=-i\int d\mathbf{p}\frac{E\pi}{4}\langle
0|\mathbf{p}\rangle \mathbf{p}p_{x}\langle
\mathbf{p}|0\rangle\sum_{n,m,n_{1},n_{2}=-\infty}^{n,m,n_{1},n_{2}=+\infty}\\
& \times\bigg\{\frac{\delta[(m-1+n+n_{1}-n_{2}){\omega}-{\omega}']}{A}\\
& +\frac{\delta[(m+1+n+n_{1}-n_{2}){\omega}-{\omega}']}{B}\bigg\}\\
&
\times\bigg[(-1)^{n_{1}+n_{2}}|u(0)|^{2}+(-1)^{n_{2}}u^{*}(0)v(0)\\
&+(-1)^{n_{1}}v^{*}(0)u(0)+|v(0)|^{2}\bigg]J_{n}J_{m}J_{n_{1}}J_{n_{2}},\\
\end{split}
\end{eqnarray}
\begin{eqnarray}
\begin{split}
&P_{0}^{1}({\omega}') = -i\int
d\mathbf{p}\frac{E\pi|Q_R|}{4Q_R}\langle 0|\mathbf{p}\rangle
\mathbf{p}p_{x}\langle
\mathbf{p}|1\rangle\sum_{n,m,n_{1},n_{2}=-\infty}^{n,m,n_{1},n_{2}=+\infty}\\
& \times\bigg\{\frac{\delta[(m-1+n+n_{1}-n_{2}){\omega}-{\omega}']}{A}\\
&+\frac{\delta[(m+1+n+n_{1}-n_{2}){\omega}-{\omega}']}{B}\bigg\}\\
&
\times\bigg[(-1)^{n_{1}+n_{2}}|u(0)|^{2}-(-1)^{n_{2}}u^{*}(0)v(0)\\
&+(-1)^{n_{1}}v^{*}(0)u(0)-|v(0)|^{2}\bigg]J_{n}J_{m}J_{n_{1}}J_{n_{2}},
\end{split}
\end{eqnarray}
\begin{eqnarray}
\begin{split}
&P_{1}^{0}({\omega}')=-i\int d\mathbf{p}\frac{E\pi
Q_R}{4|Q_R|}\langle 1|\mathbf{p}\rangle \mathbf{p}p_{x}\langle
\mathbf{p}|0\rangle\sum_{n,m,n_{1},n_{2}=-\infty}^{n,m,n_{1},n_{2}=+\infty}\\
& \times\bigg\{\frac{\delta[(m-1+n+n_{1}-n_{2}){\omega}-{\omega}']}{A}\\
&+\frac{\delta[(m+1+n+n_{1}-n_{2}){\omega}-{\omega}']}{B}\bigg\}\\
&
\times\bigg[(-1)^{n_{1}+n_{2}}|u(0)|^{2}+(-1)^{n_{2}}u^{*}(0)v(0)\\
&-(-1)^{n_{1}}v^{*}(0)u(0)-|v(0)|^{2}\bigg]J_{n}J_{m}J_{n_{1}}J_{n_{2}},
\end{split}
\end{eqnarray}
\begin{eqnarray}
\begin{split}
&P_{1}^{1}({\omega}')=-i\int d\mathbf{p}\frac{E\pi}{4}\langle
1|\mathbf{p}\rangle \mathbf{p}p_{x}\langle
\mathbf{p}|1\rangle\sum_{n,m,n_{1},n_{2}=-\infty}^{n,m,n_{1},n_{2}=+\infty}\\
& \times\bigg\{\frac{\delta[(m-1+n+n_{1}-n_{2}){\omega}-{\omega}']}{A}\\
&+\frac{\delta[(m+1+n+n_{1}-n_{2}){\omega}-{\omega}']}{B}\bigg\}\\
&
\times\bigg[(-1)^{n_{1}+n_{2}}|u(0)|^{2}-(-1)^{n_{2}}u^{*}(0)v(0)\\
&-(-1)^{n_{1}}v^{*}(0)u(0)+|v(0)|^{2}\bigg]J_{n}J_{m}J_{n_{1}}J_{n_{2}},
\end{split}
\end{eqnarray}
where$J_{n}=J_{n}(\frac{-p_{x}E}{{\omega}},\frac{E^{2}}{8{\omega}}),$$J_{m}=J_{m}(\frac{p_{x}E}{{\omega}},\frac{-E^{2}}{8{\omega}}),$$
J_{n_{1}}=J_{n_{1}}(\frac{|Q_{R}|}{{\omega}},\frac{E^{2}}{8{\omega}}),$
$J_{n_{2}}=J_{n_{2}}(\frac{|Q_{R}|}{{\omega}},\frac{E^{2}}{8{\omega}}),$
$A=(m-1+n_{1}){\omega}-\frac{\mathbf{p}^{2}}{2}+E_{0},$
$B=(m+1+n_{1}){\omega}-\frac{\mathbf{p}^{2}}{2}+E_{0}.$

From above formulae, it is easy to prove that if initially the
system is in the ground or first excited state,  only odd harmonic
can be generated and if initially the system is in the coherent
superposition state, both odd and even harmonics can be generated.

Using(16)-(19), the Fourier component of $D(t)$ reads as
\begin{eqnarray}
P({\omega}') &=&
\int\limits_{t=-\infty}^{t=+\infty}dt\left[D^{0}_{0}(t)+D^{0}_{1}(t)+D^{1}_{0}(t)+
D^{1}_{1}(t)\right]e^{-i{\omega}'t}\nonumber\\
&=&
P^{0}_{0}({\omega}')+P^{0}_{1}({\omega}')+P^{1}_{0}({\omega}')+
P^{1}_{1}({\omega}').
\end{eqnarray}

From the above analysis, one can show that no matter initially the
system being in the ground state $|0\rangle$ or first excited
state $|1\rangle$, only odd harmonics should be included in (20);
while initially the system being in the coherent superposition
state of
$\frac{\sqrt{2}}{2}|0\rangle+\frac{\sqrt{2}}{2}|1\rangle$\cite{21},
the formula (20) should include all odd and even harmonics, which
corresponds to our numerical calculation (seeing Fig.6). According
to the generalized Bessel function formula
${lim}_{u\rightarrow0}[J_{2n+1}(u,v)]=0$\cite{29}, while
$|Q_{R}|\rightarrow0$, if one or all of the values of $n_{1}$ and
$n_{2}$ are odd, then
$J_{n_{1}}(\frac{|Q_{R}|}{{\omega}},\frac{E^{2}}{8{\omega}})J_{n_{2}}(\frac{|Q_{R}|}{{\omega}},\frac{E^{2}}{8{\omega}})\rightarrow0$,
hence while $|Q_{R}|\ll 1$, in all case, odd harmonics should
primarily come from (16) and (19), and even harmonics should
primarily come from (17) and (18). However, because the initial
phases of $|0\rangle$ and $|1\rangle$ all take zero in our
numerical calculation, hence $\langle 1|\mathbf{p}\rangle\langle
\mathbf{p}|0\rangle$=$-\langle 0|\mathbf{p}\rangle\langle
\mathbf{p}|1\rangle$, and $Q_{R}/|Q_{R}|=-i$, the contributions to
even harmonics from (17) should  be counteracted by the
contributions from (18), which imply that the intensity of even
harmonics as a whole should be weaker than that of odd one, to
which the contributions from (16) and (19) shouldn't counteract
each other, and should depend on the relative phase of the ground
and first excited state. Since other contributions to even
harmonics from (16) and (19) are correlative to $n_{1}-n_{2}$
being odd, then their intensity also should depend on the value of
$|Q_{R}|$. The result also is accordant with the ultimate case of
the strict degeneracy of $|0\rangle$ and $|1\rangle$, where
$E_{1}=E_{0}$, $D=\langle0|\mathbf{p}|1\rangle$=$0$ and $Q_R$=$0$.
In the ultimate case, all even harmonic from (16)-(19) should
disappear which spells that while $R\rightarrow\infty$, the
symmetrical diatomic molecule ions should be equivalent to two
unattached atoms.
\begin{figure}[tbh]
\begin{center}
\rotatebox{0}{\resizebox *{8.5cm}{8.0cm} {\includegraphics
{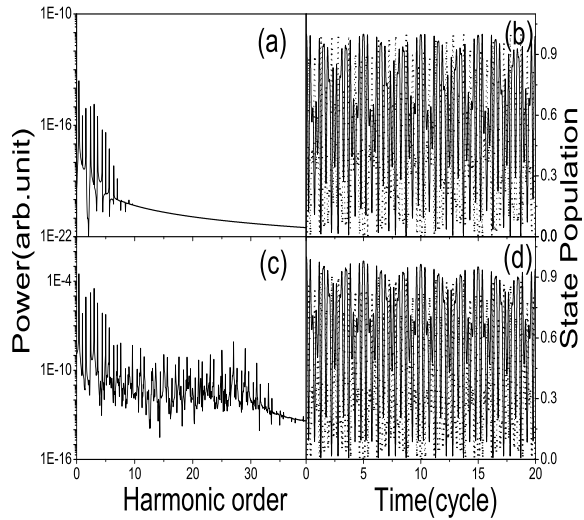}}}
\end{center}
\caption{Photon-emission spectrum of symmetrical diatomic
molecular
 ions and the population of the ground state(solid curve) and the first excited
 state(dotted curve)
 at $R=5.2a.u.$, $I=6.7\times 10^{13}W/cm^2.$ and ${\omega}=0.05642a.u.$,  initially in the ground state:
 (a)and(b) Two-level calculation; (c)and(d) $1D$ time-dependent exact calculation.} \label{fig.1}
\end{figure}
Furthermore, since in the ultimate case $R\rightarrow\infty$ and
$(E_{1}-E_{0})/{\omega}\rightarrow0$, so
${\omega}_{q}\rightarrow0$, the Floquet states theory predicts the
appearance of all odd and even harmonics while the system
initially being in the ground state\cite{13,25}, and our model
only predicts odd one in the case, it seems safe to conclude that
the even harmonics coming from the large Rabi splittings of the
odd harmonics should get weaker and weaker and disappear finally
with the increasing $R$ (seeing Fig.2 and Fig.6(a)).
\begin{figure}[tbh]
\begin{center}
\rotatebox{0}{\resizebox *{8.5cm}{8.0cm} {\includegraphics
{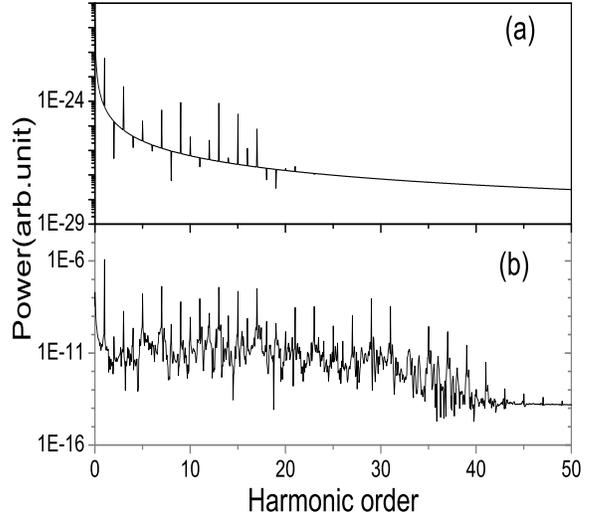}}}
\end{center}
\caption{Photon-emission spectrum of symmetrical diatomic
molecular
 ions at
 $R=16a.u.$, $I=10^{14}W/cm^2$, and ${\omega}=0.05642a.u.$, initially in the ground state:(a)
 two-level calculation;(b) $1D$ time-dependent
 exact calculation.} \label{fig.2}
\end{figure}
\section{Numerical Results}
\begin{figure}[tbh]
\begin{center}
\rotatebox{0}{\resizebox *{8.5cm}{8.0cm} {\includegraphics
{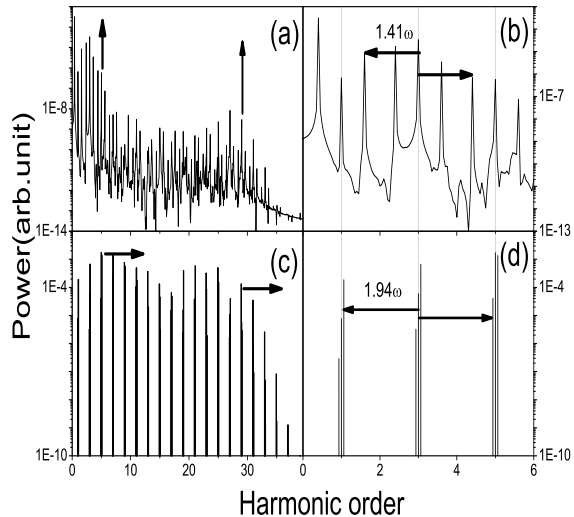}}}
\end{center}
\caption{Harmonic-generation spectra at ${\omega}=0.05642a.u.$,
$R=5.2$, $I=6.7\times 10^{13}W/cm^2$, initially in the ground
state, from
 $1D$ time-dependent exact calculation ((a) and (b)) and the formula (14)((c) and (d)):
 (a) and (c) the contour of the spectrum; (b)and (d) the corresponding fine structure of (a) and (c), respectively.
 The arrows in (a) and (c) show the molecular and the atomic like cutoffs of the transfer harmonics;
 the arrows in (b)and (d) show the symmetrical splitting around the 3th harmonic, and the numbers above the arrows
 show the splitting separation.} \label{fig.3}
\end{figure}
\begin{figure}[tbh]
\begin{center}
\rotatebox{0}{\resizebox *{8.5cm}{8.0cm} {\includegraphics
{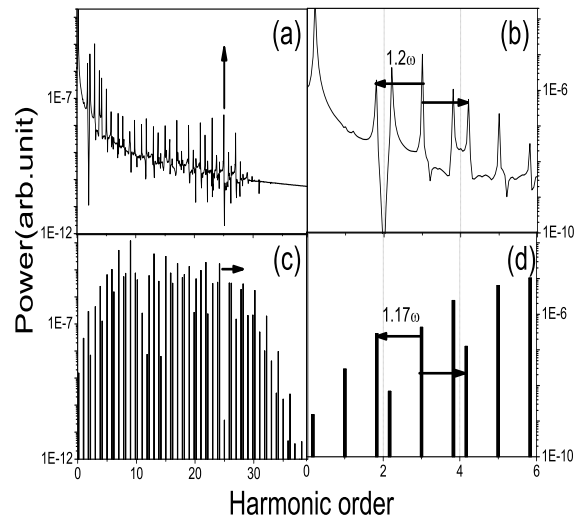}}}
\end{center}
\caption{Harmonic-generation spectra at ${\omega}=0.05642a.u.$,
$R=6a.u.$, $I=5.3\times 10^{13}W/cm^2$, initially in the ground
state, from
 $1D$ time-dependent exact calculation ((a) and (b)) and the formula (14)((c) and (d)):
 (a) and (c) the contour of the spectrum; (b)and (d) the corresponding fine structure of (a) and (c), respectively.
 The arrows in (a) and (c) show the atomiclike cutoffs of the transfer harmonics;
 the arrows in (b)and (d) show the symmetrical splitting around the 3th harmonic, and the numbers above the arrows
 show the splitting separation.} \label{fig.4}
\end{figure}
\begin{figure}[tbh]
\begin{center}
\rotatebox{0}{\resizebox *{8.5cm}{8.0cm}
{\includegraphics{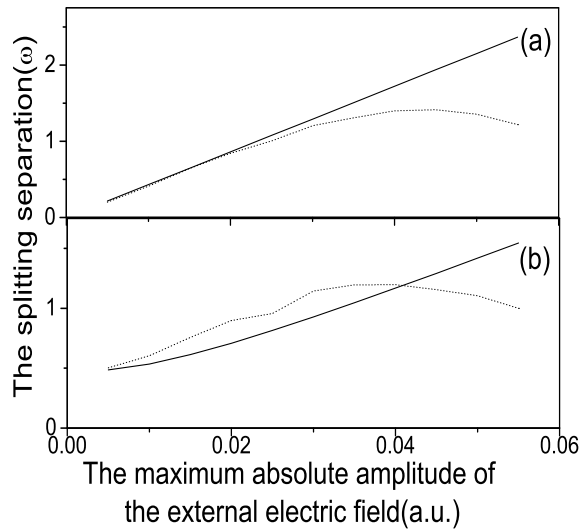}}}
\end{center}
\caption{The splitting separation(scaled by the frequency
${\omega}$) at ${\omega}=0.05642a.u.$ get from the model
prediction of $Q$(solid curve) and from $1D$ time-dependent
 exact calculation (dotted curve) VS
the maximum absolute amplitude of the external electric field
(atomic unit): (a)$R=5.2$; (b) $R=6$.
 } \label{fig.5}
\end{figure}
\begin{figure}[tbh]
\begin{center}
\rotatebox{0}{\resizebox *{8.5cm}{8.0cm} {\includegraphics
{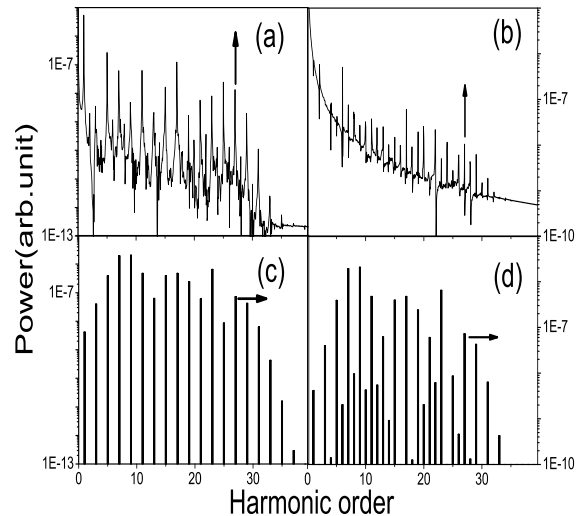}}}
\end{center}
\caption{Harmonic-generation spectra calculated from
 $1D$ time-dependent exact calculation((a) and (b)) and the formula (20) ((c) and (d)) at
 $R=8a.u.$, $I=6.7\times10^{13}W/cm^2$ and ${\omega}=0.05642a.u.$, (a) and (c)initially in the ground state;
 (b) and (d)initially in the coherent superposition state of
 the ground state and the first excited state with equally weighted
 populations.The arrows in (a) and (c) show the atomic-like cutoffs of the transfer
 harmonics.
 } \label{fig.6}
\end{figure}
\begin{figure}[tbh]
\begin{center}
\rotatebox{0}{\resizebox *{8.5cm}{8.0cm} {\includegraphics
{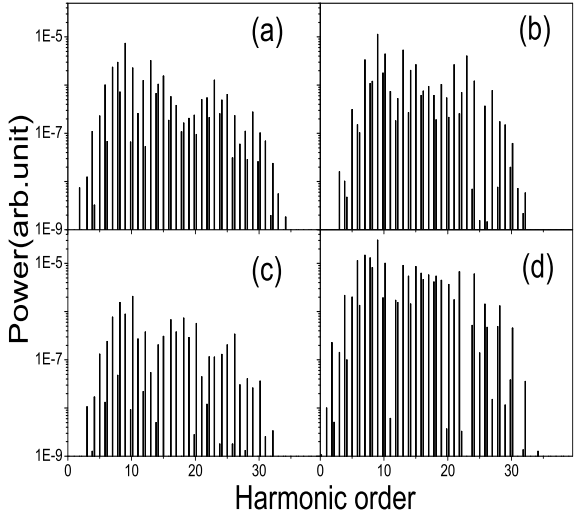}}}
\end{center}
\caption{Harmonic-generation spectra (same parameters as in
Fig.4), calculated from the formulas (10)-(13):
  (a)from (10); (b)from (11); (c)from (12); (d)from (13).
 } \label{fig.7}
\end{figure}
\begin{figure}[tbh]
\begin{center}
\rotatebox{0}{\resizebox *{8.5cm}{8.0cm} {\includegraphics
{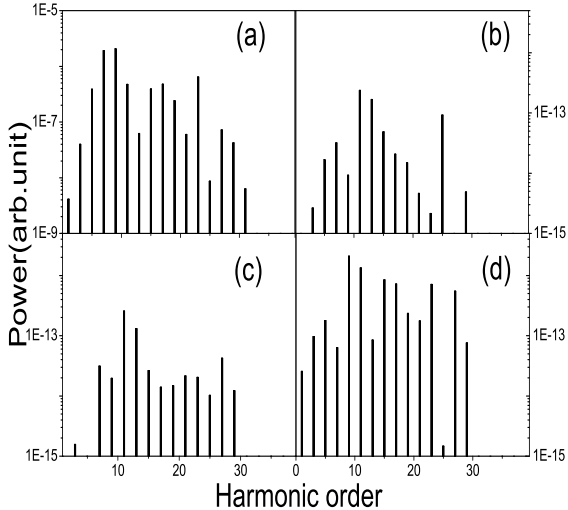}}}
\end{center}
\caption{Harmonic-generation spectra (same parameters as in
Fig.6(a)), calculated from the formulas (16)-(19):
  (a)from (16); (b)from (17); (c)from (18); (d)from (19).
 } \label{fig.8}
\end{figure}
\begin{figure}[tbh]
\begin{center}
\rotatebox{0}{\resizebox *{8.5cm}{8.0cm} {\includegraphics
{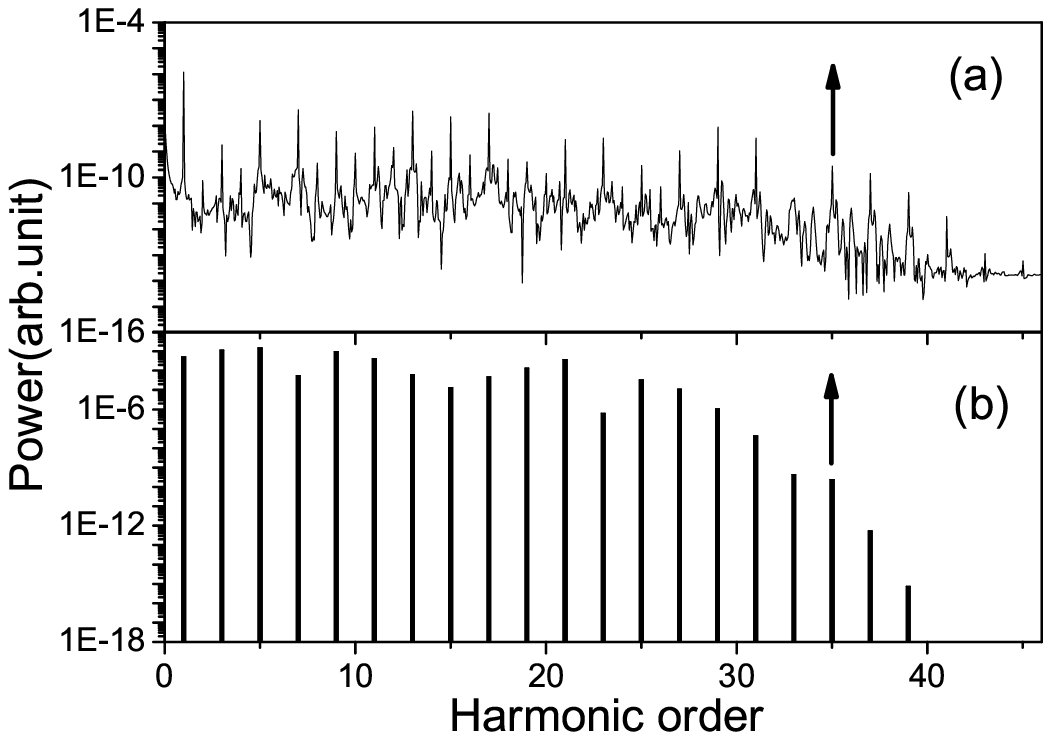}}}
\end{center}
\caption{Harmonic-generation spectra at ${\omega}=0.05642a.u.$,
$R=16$, $I=1\times
 10^{14}W/cm^2$, initially in the ground state:
  (a) $1D$ time-dependent exact calculation; (b) calculation according to the formula (20).
 The arrows in (a) and (c) show the atomic-like cutoffs of the transfer harmonics.} \label{fig.9}
\end{figure}
In this section, we apply our theory to concrete model and present
numerical results.  For convenience, we adopt an $1D$ symmetrical
diatomic molecule model. The Hamiltonian of the model molecule
studied here is
$H(t)$=$-\frac{d^{2}}{2dx^{2}}$+$\frac{z}{\sqrt{1+({x+0.5R})^{2}}}$+$\frac{z}{\sqrt{1+({x-0.5R})^{2}}}$$-xE\sin({\omega}t)$,
where $z$ is the effective charges, $R$ is the internuclear
separation, $E$ is the absolute amplitude of the external electric
field here and ${\omega}$ the frequency of the external field. In
the paper, we adopt atom unit, $\hbar=e=m_e=1$, the laser
intensity varies from $10^{13}W/cm^{2}$ to
$10^{14}W/cm^{2}$($0.017a.u.-0.055a.u.$), the internuclear
separation varies from intermediate $R$ ($5.2a.u.,$ and $6a.u.$)
to large $R$ ($16a.u.$), the wavelength adopted is
$\lambda$=$800nm$, accordingly ${\omega}=0.05642a.u.$ and the
laser pulse contains 20 optical cycles. Numerically the above
schr\"odinger equation can be solved by operator-splitting
method\cite{22}.

 We begin the
calculation by assuming the electron initially is in the ground
state. But Ivanov {\it et al} show that one should consider at
least two wave packets excited to the electronic surfaces in their
program\cite{32}, which means initially the system should be in
the coherent superposition state\cite{21}. So we also perform the
calculations from the coherent superposition state of the ground
state and the first excited state with equally weighted
populations.
\subsection{Failure of the two-level modle in producing the whole harmonic spectrum}
 When $R=5.2a.u.$, the ionization potential of ground state is
$E_{0}=-0.89830a.u.$ and first excited state is $E_{1}
=-0.84085a.u.$. The energy separation between $E_{0}$ and $E_{1}$
is $\triangle E=0.05745a.u.$, which is in the
near-(one-photon)-resonance region. Fig.1 shows the spectrum from
$1D$ time-dependent calculation for the model molecule at
$R=5.2a.u.$. In Fig.1, each odd harmonic peak is accompanied
symmetrically by two strong sidebands. The energy separations
between the neighboring sidebands all have the same value, which
can be explained in a dressed molecule state picture\cite{33}.

Comparison between Fig.1 (a) and Fig.1(c) shows that while the
external field becomes stronger and although the ground states and
the first excited state depletion in Fig.1(c) is only $0.03$,
there is very large difference between $1D$ time-dependent exact
calculation and two-level calculation. So the continuum state
effects on spectrum must be considered in this case. Otherwise, an
obvious characteristic, different from atom, for the population of
molecule states is the strong-coupling of the ground state and the
first excited state. It shows that it is necessary and reasonable
to consider the first excited state effect on harmonic generation
to explain the spectrum from symmetrical diatomic molecular ions.

While $R=16a.u.$, the ionization potentials of the ground state
and the first excited state are $E_{0}=-0.732574a.u.$ and
$E_{1}=-0.7325723a.u.$, respectively. The energy separation is
$\triangle E=1.7\times 10^{-6}a.u.$, which is in the
strong-coupling region. When $I=10^{14}W/cm^2$, and the depletion
of ground states and the first excited state is only $0.025$.
Fig.2(b) exhibits harmonic-generation efficiencies of at least
four orders of magnitude greater than that of Fig.2(a). It is
worth it to be noted that both odd and even harmonic peaks appear
although the even harmonic is weak in Fig.2, which accords with
our above analysis that with the increasing internuclear
separation $R$, the Rabi splitting of the odd harmonics should
converge towards even harmonics and their intensity should become
weaker and weaker. Again the calculation result also shows the
strong-coupling of the ground state and the first excited state.
\subsection{Harmonic spectrum  calculated from our model}
Fig.3(a) and (c) show the HOHG spectra from the $1D$
time-dependent calculation and the numerical calculation of the
formula (14) with $R$=$5.2a.u.$ and assuming the system initially
being in the ground state. Fig.3(b) and Fig. 3(d) show the fine
structure of Fig.1(a) and Fig.3(c), respectively. The splitting
separation is $Q=1.41w$ in Fig.3(b) and $1.94w$ in Fig.3(d). The
discrepancy is due to invalidation of rotating-wave approximation
in intense field. A plateau in low-order region is also
identifiable in the spectrum in Fig.3(c) with a cutoff at the 5th
harmonic order, which agrees with Fig.3(a). This corresponds to
the maximum energy acquired by the electron in the two-level
system in the presence of the field\cite{26}. Furthermore, a
second plateau up to the $25th$ harmonic order which is associated
with the ground-continuum coupling (referred to as the atomic
plateau) can be identified. The cutoff position of the atomic
plateau can be well explained by a semiclassical model\cite{28}.

Fig.4 shows the HOHG spectra from $1D$ time-dependent numerical
calculation and (14) with $R$=$6a.u.$ and $E_{1}-E_{0}=0.0304a.u.$
The total depletion of the ground states and the first excited
state is $0.34$. The splitting separations
 is $Q=1.2{\omega}$ in Fig.4(b) and $1.17{\omega}$ in Fig.4(d). The position
of cutoff in Fig.4(c) is at the $24th$ harmonic order that agrees
with the order $n=(3.17U_{p}+I_{p})/{\omega}$, the cutoff law
predicted by the semiclassical mechanism of HOHG.

Fig5 shows the comparison of the splitting separation(scaled by
the frequency ${\omega}$ with ${\omega}=0.05642a.u.$) between the
model prediction of $Q$=$\sqrt{|Q^{2}_{R}|+\xi^{2}}$(solid curve)
and the $1D$ time-dependent
 exact calculation (dotted curve) at $R$=$5.2a.u.$(Fig5(a)) and $R$=$6a.u.$(Fig5(b)).
 It is obvious that in the near resonance
 case(Fig5(a)), good agreement is obtained while the field is
 weak. However while the field becomes stronger, the difference
 between the theory one and the numerical one seems smaller in
 Fig5(b) than in Fig5(a) with the same field intensity.
\subsection{Influence of  initial condition and internuclear distance}
At intermediate $R$, such as $R$=$5.2a.u.$ or $R$=$6a.u.$, no
matter initially the system is in the ground state or the first
excited state, or the coherent superposition state, no distinct
difference is found in the spectra from numerical calculations.
However at large $R$, the case is rather different. Fig.6 shows
the comparison of harmonic generation spectrum calculated from
different initial condition at $R$=$8a.u.$, where
$E_{1}-E_{0}=0.0052a.u.$ and $Q=0.95{\omega}$. In Fig.6(b), where
the system initially is in the coherent superposition state, even
harmonics appear although they are weaker than the odd harmonics,
however in Fig6(a), where the system initially is in the ground
state, even harmonics hardly are distinguishable. The result
validates our analysis for (20), which predicts that at large $R$
if the system initially is in the ground state, only odd harmonic
can be produced , and if it is in the coherent superposition
state, even harmonics also should be emitted although their
strength should be weaker than the odd ones. If initially the
system is in the excited state, the case is analogous with the
ground state. Other calculations at $R$=$16a.u.$ from different
initial conditions also show similar result. The position of
cutoff in Fig.6(c) is at the $23th$ harmonic order, which agrees
with the order $n=(3.17U_{p}+I_{p})/{\omega}$.

At intermediate $R$, with the same parameters as in Fig4, Fig.7
shows the comparison of contributions to harmonic from the four
different terms (10)-(13). It can be seen that all of them give
equivalent  contribution to HOHG in this case.

At large $R$, with the same parameters as in Fig6(a), Fig.8 shows
the comparison of contributions to harmonic from the four
different terms (16)-(19) if initially the system is in the ground
state. It is the term of
$|0\rangle\rightarrow|\mathbf{p}\rangle\rightarrow|0\rangle$ that
contributes mostly to HOHG in the case. Similarly, the calculation
also shows that if initially the system is in the first excited
state, the primary contribution to HOHG should come from
$|1\rangle\rightarrow|\mathbf{p}\rangle\rightarrow|1\rangle$.

\subsection{Cutoff law at large $R$}
Fig.9 shows the spectra calculated by our analytic theory and $1D$
numerical simulation with $R$=$16.a.u.$. The main difference is
that there are weak even harmonic peaks in Fig.9(a) but no even
harmonic peaks in Fig.9(a). This is because $E_{1}-E_{0}=0$ has
been adopted in the analytic formulation but actually these two
states are not completely degenerated. The position of cutoff in
Fig.9(b) is at the $35th$ harmonic order, which agrees with the
order $n=(5.33U_{p}+I_{p})/{\omega}$ that is obtained by the
semiclassical calculation considering complex trajectories. This
is consistent with the finding that at very large $R$, the
molecule has different cutoff law \cite{2,13}.

\section{Conclusions}
In summary, with considering the continuum states and ionization,
the photon-emission spectra of the symmetric diatomic molecule
ions in intense laser fields are  thoroughly investigated and the
role of the CR states is identified.

In the near-resonance region ($(E_{1}-E_{0})/{\omega}\simeq 1$),
it is shown that the multiple Mollow triplet structure can be
described by an analytic theory based on rotating-wave
approximation and the Rabi oscillation between the ground state
and the first excited state induces the symmetrical splittings of
odd harmonics. The splitting separation in weak field can be
approximately denoted by
$Q$=$\sqrt{|ED|^{2}+(E_{1}-E_{0}-{\omega})^{2}}$.

In the strong coupling region ($(E_{1}-E_{0})/{\omega}\simeq 0$),
all of the symmetrical splitting sidebands of odd harmonics
gradually converge to even harmonics and their amplitudes get
weaker and weaker with the increasing internuclear separation $R$.
This is explained by our theory  in the ultimate condition of
$(E_1-E_0)/{\omega}\simeq 0$($R\rightarrow\infty$). An interesting
phenomenon is that in this case if initially the system is in the
coherent superposition of the ground state and the first excited
states, there are stronger even harmonics emitted than only in the
ground or first excited state, and their intensity depends on the
value of $|Q_R|=|ED|$.

In all above cases, the molecular plateau and the atomic plateau
are identifiable in the analytical model and the cutoff positions
are accordant with the numerical simulation and semiclassical
calculation.

We also find that, for the symmetrical diatomic molecule, the
dipole movement includes four terms, those are
$|0\rangle\rightarrow|\mathbf{p}\rangle\rightarrow|0\rangle$,
$|1\rangle\rightarrow|\mathbf{p}\rangle\rightarrow|0\rangle$,
$|0\rangle\rightarrow|\mathbf{p}\rangle\rightarrow|1\rangle$ and
$|1\rangle\rightarrow|\mathbf{p}\rangle\rightarrow|1\rangle$,
respectively, denoting the transitions from the ground state or
the first excited state to the continuum state, then back to the
ground state or the first excited state. It is more complicated
than atom case because of the existence of the $CR$ states. At
intermediate $R$, each process contributes equivalently to
molecular HOHG. However, at very large $R$, their contributions
depend on the initial condition.

\section{Acknowledgements}
This work is supported National Hi-Tech ICF Committee of China
under Grant No. 2004AA84ts08 and the National Natural Science
Foundations of China under Grant No. 10574019.

\end{document}